\documentclass[twocolumn,amsmath,nofootinbib,amssymb,superscriptaddress,longbibliography]{revtex4-1}
\usepackage{natbib}
\usepackage{graphicx}
\usepackage{color}
\usepackage[colorlinks, linkcolor=blue]{hyperref}
\usepackage{amssymb}
\usepackage{gensymb}
\usepackage{float}
\usepackage{amsmath}
\usepackage{tabularx,graphicx}
\usepackage{epstopdf}
\usepackage{latexsym}
\usepackage{color, colortbl}
\usepackage{psfrag}
\usepackage{bbm}
\usepackage{bm}
\usepackage{blindtext}
\usepackage{dsfont}
\usepackage{slashed}
\usepackage{multirow}
\usepackage{appendix}
\usepackage{ragged2e}
\usepackage{bbold}
\usepackage{braket,mleftright}

\def \be{\begin{equation}}
\def \ee{\end{equation}}
\def \ba{\begin{array}}
\def \ea{\end{array}}
\def \bea{\begin{eqnarray}}
\def \eea{\end{eqnarray}}

\def \ve {\varepsilon}
\renewcommand{\vec}[1]{\boldsymbol{#1}}

\def \bL{{\bf L}}

\def \br{{\bf r}}

\def \bz{{\bf z}}
\def \by{{\bf y}}
\def \bx{{\bf x}}
\def \bn{{\bf n}}

\def \e{{\epsilon}}

\def \a{{\alpha}}
\def \t{{\theta}}
\def \b{{\beta}}
\def \g{{\gamma}}
\def \D{{\Delta}}

\def \e{{\epsilon}}
\def \ve{{\varepsilon}}

\def \G{{\Gamma}}
\def \z{{\zeta}}

\newcommand{\ehat}{\hat {\bf e}}

\def \ba{\begin{align*}}
\def \ea{\end{align*}}

\newcounter{indice}

\def \mrm{\mathrm}

\def \bs{\boldsymbol}
\def \mc{\mathcal}

\date{~\today}

\begin{document}
\title{Flat band physics in the charge-density wave state of $1T$-TaS$_2$}

\author{Amir Dalal}
\affiliation{Department of Physics, Bar-Ilan University, 52900, Ramat Gan, Israel}
\affiliation{Center for Quantum Entanglement Science and Technology, Bar-Ilan University, 52900, Ramat Gan Israel}

\author{Jonathan Ruhman}
\affiliation{Department of Physics, Bar-Ilan University, 52900, Ramat Gan, Israel}
\affiliation{Center for Quantum Entanglement Science and Technology, Bar-Ilan University, 52900, Ramat Gan Israel}

\author{J\"orn W. F. Venderbos}
\affiliation{Department of Physics, Drexel University, Philadelphia, PA 19104, USA}
\affiliation{Department of Materials Science \& Engineering,
Drexel University, Philadelphia, Pennsylvania 19104, USA}

\begin{abstract}
1$T$-TaS$_2$ is the only insulating transition-metal dichalcogenide (TMD) with an odd number of electrons per unit cell. This insulating state is non-magnetic, making it a potential spin-liquid candidate. The unusual electronic behavior arises from a naturally occurring nearly flat mini-band, where the properties of the strongly correlated states are significantly influenced by the microscopic starting point, necessitating a detailed and careful investigation.
We revisit the electronic band structure of 1$T$-TaS$_2$, starting with the tight-binding model without CDW order. Symmetry dictates the nature of spin-orbit coupling (SOC), which, unlike in the 2H TMD structure, allows for strong off-diagonal ``spin-flip'' terms as well as Ising SOC. Incorporating the CDW phase, we construct a 78$\times$78 tight-binding model to analyze the band structure as a function of various parameters. Our findings show that an isolated flat band is a robust feature of this model. Depending on parameters such as SOC strength and symmetry-allowed orbital splittings, the flat band can exhibit non-trivial topological classifications. These results have significant implications for the strongly correlated physics emerging from interacting electrons in the half-filled or doped flat band.
% 1$T$-TaS$_2$ is the only insulating transition-metal dichalcogenide (TMD) with odd number of electrons per unit cell.
% This unusual  behavior owes to the natural formation of a nearly flat band following a charge-density wave (CDW) transition. The properties of the strongly correlated states of this flat band are strongly influenced by the microscopic model and therefore call for a careful study of these states.
% We revisit the question about the electronic band structure of 1$T$-TaS$_2$. We start with the tight-binding model without the CDW order. Symmetry dictates the nature of spin-orbit coupling (SOC), which differs from that in the 2H structure by allowing off-diagonal "spin-flip" terms as well as the Ising SOC. We then add the CDW phase and construct a 78$\times$78 tight-binding model and analyze its band structure as a function of parameters. We find that an isolated flat band is a relatively robust of this model. Depending on parameters such as SOC strength and symmetry-allowed orbital splittings the flat band can also develop non-trivial topological classification. Our results have dire implications for the strongly correlated physics emerging from interacting electrons in the half-filled or doped flat band.
\end{abstract}

\maketitle

%\tableofcontents

\section{Introduction}
Flat dispersion bands have the  propensity to form a wide range of interaction-driven quantum phenomena. For this reason they are important to fundamental science as well as quantum technological applications~\cite{checkelsky2024flat}.
Such bands may emerge due to interactions or strong orbital hybridization~\cite{Zhitomirsky2004,bistritzer2011moire}. The latter case has recently drawn a significant amount of interest due the dominance of quantum geometrical effects~\cite{torma2023essay}.
Examples of interaction driven quantum phases resulting from flat bands include spin-liquid states, the fractional quantum Hall effect~\cite{parameswaran2013fractional}, fractional Chern insulators~\cite{bergholtz2013topological}, unconventional superconductivity and various symmetry breaking states~\cite{balents2020superconductivity}

An interesting realization of a nearly flat band naturally occurs in the transition metal dichalcogenide (TMD) $1T$-TaS$_2$~\cite{smith1985band}. At a temperature of $T~\sim~200$ K this material undergoes a transition to a commensurate charge density wave (CDW) state, leading to a redistribution of the electronic charge and a large lattice distortion~\cite{wilson1975charge}. In this CDW state, clusters of 13 Ta atoms form a triangular lattice of stars-of-David (SoD), as shown in Fig.~\ref{fig:CDWLattice}. Consequently, the band structure is dramatically reconstructed as a result of the CDW distortion, giving rise to the \emph{spontaneous} formation of a nearly flat and half-filled band. Notably, the band is isolated from all other bands by gaps of the order of 100 meV.

At low temperature 1$T$-TaS$_2$ is found to be a non-magnetic insulator. It is commonly accepted that the mechanism behind the insulating behavior is Mott physics driven by Coulomb repulsion~\cite{di1977low}. No sign of magnetic ordering has been detected down to lowest cryogenic temperatures~\cite{ribak2017gapless}, which makes this system a natural spin-liquid candidate~\cite{law20171t}. Support for possible spin liquid physics was offered by STM experiments on thin layers of 1$T$-TaSe$_2$~\cite{regan2020mott}, which found incommensurate peaks in the quasi-particle interference patterns consistent with a spinon Fermi surface (SFS). However, theoretical and experimental studies propose that the bulk insulating behavior stems form interlaer dimerization (see for example Ref.~\cite{ritschel2018stacking}).

From the theoretical perspective, efforts to understand the intriguing electronic and magnetic properties observed in 1$T$-TaS$_2$ have focused on two separate but related aspects of the problem. The first concerns the electronic properties of the flat band~\cite{smith1985band}. First-principles electronic structure calculations suggested that spin-orbit coupling is crucial for the formation of the flat band~\cite{rossnagel2006spin}, but more recent work has shown that an isolated flat band can emerge even when spin-orbit coupling is ignored~\cite{YuXiangLongElectronic,LarsonEffects}. In particular, a recent study revealed that an insightful tight-binding description of the relevant Ta-derived bands can be obtained by choosing an orbital basis tied to the local near-octahedral environment of the TaS$_6$ clusters~\cite{cheng2021understanding}. In this basis, in which the active bands are formed from $t_{2g}$ states, the band dispersion obtained from first-principles calculations can be approximated by restricting to intra-orbital hopping, which provides a useful idealized limit for understanding the emergence and structure of the flat band.

While the electronic structure of the flat band is generally addressed within single-particle band theory, either through first-principles calculations or effective Hamiltonian modeling, gaining insight into the observed nonmagnetic insulator at low temperatures requires including electron correlations. This aspect of the problem (i.e., the Mott physics) has therefore been addressed within the framework of an effective multi-orbital Hubbard model---a common approach in the realm of strongly correlated electron materials~\cite{law20171t}. Based on this description an XXZ spin Hamiltonian has been proposed, based on which the magnetic ordering and spin liquid phenomena can be studied~\cite{he2018spinon}.

The construction of an effective Hubbard model and resulting low-energy spin-spin model relies on an appropriate microscopic starting point. In the case of 1$T$-TaS$_2$, this means obtaining an accurate model for the flat band, from which a suitable spin model can then be derived. With this motivation in mind, we construct a microscopic tight-binding model for 1$T$-TaS$_2$ and revisit the precise nature of the flat band in 1$T$-TaS$_2$. We do so by approaching the band reconstruction in the CDW state and the resulting emergence of an isolated flat band from a different perspective compared to previous work. In our approach the CDW state is viewed as a superlattice of weakly coupled SoD clusters, such that the minibandstructure in the CDW state can be interpreted as a collection of nearly flat dispersive bands derived from the energy levels of the individual clusters. Viewed in this light, the nature of the flat band can be traced back to the molecular orbitals of the SoD clusters, which we systematically construct. We will show that this gives important insight into the flat band physics and leads to a unified understanding of previous descriptions of the band reconstructions. We highlight in particular the role played by spin-orbit coupling.

Exploiting the paradigm of weakly coupled clusters, we examine and discuss three possible flat band scenarios, which are sharply distinguished by the symmetry and structure of the cluster levels. The realization of these scenarios depends on the relative strength of parameters of the model which affect the ordering of energy levels of a single SoD cluster, such as crystal field splitting, hopping, and spin-orbit coupling. We show that it is possible, at least in principle, to achieve a level crossing of cluster states with opposite parity, which can give rise to topological minibands once the clusters are coupled to form a superlattice.

Our analysis of the CDW state as coupled clusters thus provides a unifying framework for addressing the single-particle flat band physics of 1$T$-TaS$_2$ and also lays the foundation for microscopically grounded descriptions of the Mott state.

%examining the band reconstruction in the CDW state from the perspective of weakly coupled SoD clusters. In this approach the SoD clusters are considered building blocks of the CDW superlattice, such that the minibandstructure of the CDW state can be viewed as nearly flat dispersive bands derived from the energy levels of the individual clusters.

%with nearly flat dispersion due to the weak coupling between the clusters.

\section{Tight binding model for $1T$-TaS$_2$ \label{sec:TB-model}}

We begin by introducing a microscopic tight-binding model for $1T$-TaS$_2$. We proceed in three steps. First, we consider a tight-binding model for an undistorted $1T$-TaS$_2$ monolayer without spin-orbit coupling, which serves as the root model for the next steps. We then include the effect of spin-orbit coupling and comment on the differences with the $1H$ $D_{3h}$ TMDs. In the final step, we include the effects of the CDW by periodically alternating the parameters in the tight-binding model.

\begin{figure}
    \centering
         \includegraphics[width=\linewidth]{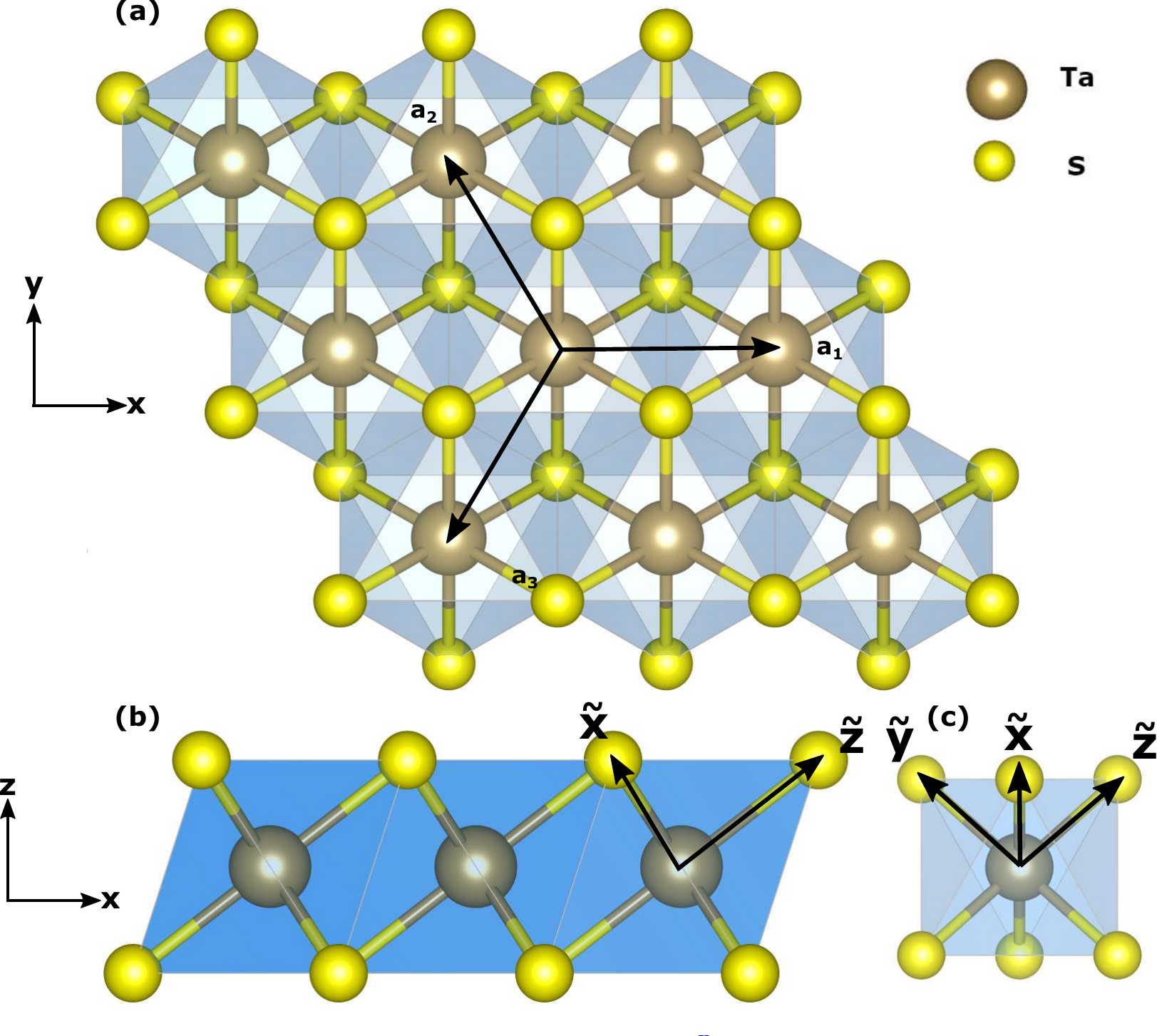}
         \caption{The crystal structure of $1T$-TaS$_2$. Top and side views are shown in panels (a),(b) respectively. A single octahedron with the corresponding aligned axes is shown in panel (c).  }
         \label{fig:CrystalStructure}
\end{figure}

\subsection{Tight-binding model in the absence of SOC and the CDW}
\label{ssec:TB-no-SOC}

The crystal structure of a $1T$-TaS$_2$ monolayer is shown in Fig.~\ref{fig:CrystalStructure}. The tantalum atoms form a triangular lattice that sits between two triangular lattices of sulfur atoms. Importantly, the top and bottom triangular sulfur layers are shifted, such that each Ta atom is surrounded by six S atoms forming a trigonally distorted octahedron (see panels b and c of Fig.~\ref{fig:CrystalStructure}). The resulting structure has trigonal $D_{3d}$ point group symmetry, generated by the threefold rotation $\mathcal C_{3z}$, the twofold rotation $\mathcal C_{2x}$, and the inversion $\mathcal I$.

The atomic orbitals relevant for the electronic structure predominantly belong to the $5d$ multiplet. To understand the effect of crystal field splittings on the $5d$ states it is instructive to consider the (distorted) octahedral environment of the Ta atoms. If the distortion away from perfect octadehral coordination is neglected, then the locally cubic environment splits the $d$ states into a $t_{2g}$ and an $e_g$ multiplet. The three $t_{2g}$ states $(d_{\tilde y \tilde z},d_{\tilde z \tilde x},d_{\tilde x \tilde y})$ are lower in energy and constitute the relevant valence electron manifold. To denote the $t_{2g}$ states here we have used a coordinate system $(\tilde x,\tilde y,\tilde z)$ aligned with the octahedron, which is rotated with respect to the crystallographic coordinate system $( x,y, z)$, as shown in Fig.~\ref{fig:CrystalStructure}(b) and (c). The usefulness of a description in terms of the octahedral coordinate system was pointed out in Ref.~\onlinecite{cheng2021understanding}, which showed in particular that such description provides a basis for considering insightful idealized limiting cases (e.g., neglecting interorbital hopping).

The crystal field generated by the trigonal distortion of the octahedra splits the $t_{2g}$ states further into a single non-degenerate level and a degenerate doublet. In the crystallographic coordinate system, which is the natural choice when distortions are properly accounted for, these states are given by $d_{z^2}$ and $(d_{x^2-y^2},d_{xy})$ (with a symmetry-allowed admixture of $d_{xz}$ and $d_{yz}$). These states transform as the $A_{1g}$ and $E_g$ representations of the $D_{3d}$ point group. To capture the electronic properties of $1T$-TaS$_2$ close to the Fermi energy, we consider a tight-binding model for these three orbitals. We choose a basis of states given by $\{|d_{+}\rangle,|d_{0}\rangle,|d_{-}\rangle\}$, where $d_0 \equiv d_{z^2}$ and $d_\pm \equiv i d_{x^2-y^2}\pm  d_{xy}$. (We emphasize that the $d_{\pm}$ states also have support on $d_{xz} \pm i d_{yz}$, see Appendix~\ref{app:basis}.) In the chosen basis, the tight-binding Hamiltonian takes the form
\begin{align}
    H_0(\mathbf{k})=\begin{pmatrix}
    \varepsilon_{1}+t_2\gamma_\mathbf{k} & t_{3}\lambda_\mathbf{k} & t_{4}\lambda^*_\mathbf{k}\\
    c.c & \varepsilon_0+t_1\gamma_\mathbf{k} & t_{3}\lambda_\mathbf{k}\\
    c.c & c.c & \varepsilon_{1}+t_2\gamma_\mathbf{k}
    \end{pmatrix},
    \label{eq:TrigH}
\end{align}
where we have included all symmetry-allowed couplings up to nearest neighbor hopping. In particular, here $\varepsilon_{0,1}$ are the on-site energies of the $d_0$ and $d_\pm$ states, such that the crystal field splitting is $\D = \varepsilon_{1}-\varepsilon_{0}$, and $t_{1,2}$ are the intra-orbital nearest neighbor hopping amplitudes. The interorbital hopping amplitudes are given by $t_3$ and $t_4$, and the functions $\gamma_\mathbf{k}$ and $\lambda_\mathbf{k}$ are defined as
\begin{equation}
\gamma_\mathbf{k} = c_1 + c_2 + c_3, \quad \lambda_\mathbf{k}  =c_1  +\omega c_2  +\omega^2 c_3,  \label{eq:gamma-lambda}
\end{equation}
where $c_i  \equiv {\cos\left({\mathbf{k}\cdot\mathbf{a}_i}\right)}$ and $\omega\equiv e^{i2\pi/3}$. Here $\mathbf{a}_i$ are the primitive lattice vectors $\vec{a}_i=\cos{\phi_i}\hat{x}+\sin{\phi_i}\hat{y}$ with $\phi_i=2\pi(i-1)/3$.

\begin{figure}

    \centering
         \includegraphics[width=\linewidth]{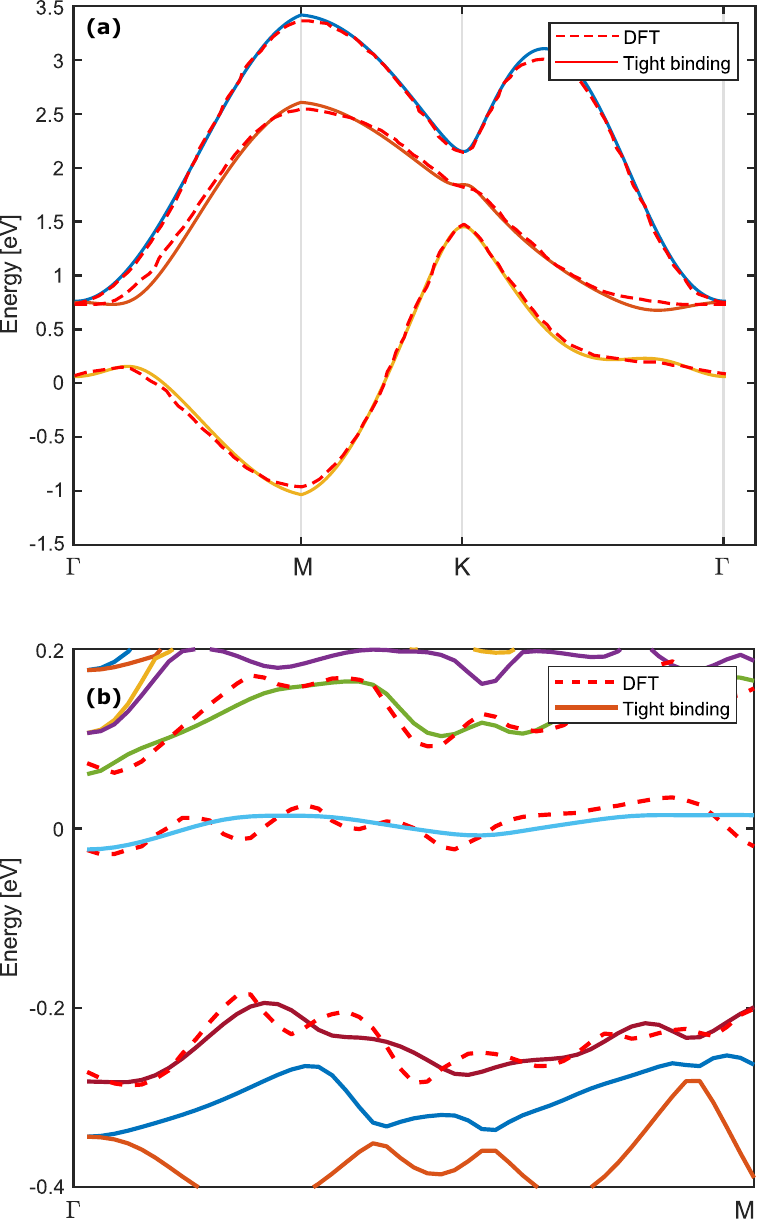}
         \caption{Fit of the tight binding parameters to the DFT result \cite{rossnagel2006spin}. (a). The dispersion of $1T$-TaS$_2$ without CDW effects. Obtained by diagonalizing the Hamiltonian $H=H_0+H_{SOC}(L_2,\theta=0)$ and fit to DFT data. (b). The dispersion of $1T$-Ta$S_2$ in the CDW phase. Obtained by diagonalizing Eq \eqref{eq:CDWH}. In both figures, the solid lines are the bands obtained from the best fit corresponding to the parameters in Table~\ref{tab:table2}. The dashed lines are the bands from the DFT data. }
         \label{fig:FitDFTBoth}
\end{figure}

The values of $t_1,\ldots t_4$, $\e_0$ and $\e_1$, which obtain the best fit with DFT calculations~\cite{rossnagel2006spin} are listed in Table.~\ref{tab:table2}.
The resulting bands and their comparison to DFT are presented in Fig.~\ref{fig:FitDFTBoth}.(a).

\subsection{Spin-Orbit Coupling \label{ssec:TB-SOC}}

The next step is to include the effect of spin-orbit coupling (SOC). The form of the spin-orbit interaction depends in a crucial way on the nature of the orbitals involved, in particular the nature of the $| d_\pm \rangle$ states, as we will explain below~\cite{liu2013three-band,de2018tuning}. In the case of $1T$-TaS$_2$ the appropriate form of the spin-orbit coupling Hamiltonian is given by
\be
H_{\text{SO}}= \lambda_0 \cos\theta L_z \sigma_z + \lambda_0 \sin\theta ( L_x \sigma_x+L_y \sigma_y), \label{eq:H_SO}
\ee
where $(\sigma_x,\sigma_y,\sigma_z)$ are the three spin Pauli matrices and $(L_x,L_y,L_z)$ are spin angular momentum matrices in the space of $d$-orbitals. Here we have introduced an overall energy scale associated with spin-orbit coupling, given by $\lambda_0$, and an angle $\theta$, which determines the relative weight of the spin-preserving Ising and spin-flip terms. As such, $\theta$ interpolates between the Ising limit ($\theta=0$) and the isotropic limit ($\theta=\pi/4$).

Many previous studies of TMD materials have considered electronic models that only include the Ising spin-orbit coupling. This is appropriate for the $1H$ materials with $D_{3h}$ symmetry, since in these systems the $(d_{x^2-y^2},d_{xy})$ orbitals are symmetry-distinct from the $(d_{xz},d_{yz})$ orbitals, forbidding these states to mix. Projecting the atomic spin-orbit coupling of the $L=2$ $d$-states with a $S=1/2$ spin into the subspace of the $(d_{x^2-y^2},d_{xy})$ and $d_z^2$ orbitals (i.e., the relevant valence and conduction states), yields the Ising term only; the spin-flip terms vanish. This is markedly different in the $1T$ structures, where the $(d_{x^2-y^2},d_{xy})$ and $(d_{xz},d_{yz})$ orbitals can mix, forming the $d_\pm$ doublet introduced above. Due to the admixture of $(d_{xz},d_{yz})$ the spin-flips do no strictly vanish, giving rise to the more general form \eqref{eq:H_SO}. The angle $\theta$ is therefore related to the orbital content of the $d_\pm$ states. More details are provided in Appendices~\ref{app:basis} and \ref{app:soc}.

In the specific context of $1T$-TaS$_2$ only the Ising spin-orbit coupling was previously considered~\cite{liu2013three-band,he2018spinon}. Its effect on the electronic states is to split the orbital doublet $|d_\pm \rangle$ into two Kramers doublets, which constitutes one possible scenario for realizing a single twofold Kramers degenerate flat band in the CDW state. We return to this
point in Sec.~\ref{sec:CDW_clusters} when we discuss the nature of the flat band. There are two compelling reasons to consider the more general form of the spin-orbit coupling Hamiltonian given by \eqref{eq:H_SO}. First, including only the Ising term leaves an unphysical continuous spin rotation symmetry around the $z$ axis, which, for instance, has consequences for the form of the effective low-energy spin Hamiltonian derived from the microscopic multi-orbital model once interactions are included. We briefly discuss this in Sec.~\ref{sec:discuss}. Second, the more general form gives access to the isotropic limit, given by $\theta=\pi/4$, which is useful for gaining insight into electronic structure of $1T$-TaS$_2$. The isotropic limit corresponds to perfect octahedral coordination of the S atoms, in which case spin-orbit coupling can simply be understood as arising from three degenerate $t_{2g}$ states $(d_{\tilde y \tilde z},d_{\tilde z \tilde x},d_{\tilde x \tilde y})$ in the octahedral coordinate frame (see Appendices~\ref{app:soc}). When the trigonal distortion is weak, which is a valid assumption in $1T$-TaS$_2$, the isotropic limit is expected to provide a good
approximation to the value of $\theta$.

% representation of the effect of spin-orbit coupling.

%\subsection{Liתst of all parameters}
\begin{table*}
\caption{\label{tab:table2}
List of parameters used in the tight binding model, with and without CDW}
\begin{ruledtabular}
\begin{tabular}{ p{3cm} p{1cm} p{1cm} p{6.5cm} p{3cm} }
 Lattice parameters  & Value & Units & Description & Reference \\
\hline
a & 3.36 & \AA & Lattice constant & Ref.~[\onlinecite{wilson1975charge}]\\
$\omega$ & $e^{i2\pi/3}$  & - & Trigonal phase factor & \\
$\varepsilon_0$ & 1.48 & eV & On-site energy of $d_0$ orbital & Fitted to DFT\cite{rossnagel2006spin} \\
$\varepsilon_1$ & 1.34 & eV & On-site energy of $d_{\pm}$ orbitals & Fitted to DFT\cite{rossnagel2006spin} \\
$t_0$ & -0.1 & eV & $d_0$-$d_0$ hopping parameter     & Fitted to DFT\cite{rossnagel2006spin}\\
$t_1$ & -0.15 & eV & $d_\pm$-$d_\pm$ hopping parameter & Fitted to DFT\cite{rossnagel2006spin}\\
$t_3$ & -0.28 & eV & $d_0$-$d_\pm$ hopping parameter   &  Fitted to DFT\cite{rossnagel2006spin}\\
$t_4$ & 0.43 & eV & $d_\pm$-$d_\mp$ hopping parameter &  Fitted to DFT\cite{rossnagel2006spin}\\
$\lambda_0$ & 0.640  & eV & SOC strength & Phenomenological Parameter (value is best fit to Ref.\cite{rossnagel2006spin}) \\
$\theta$ & --  & - & {Relative strength of spin-flip and spin conserving SOC terms  ($\theta=0$ is the Ising limit and $\theta=\pi/4$ is the isotropic limit)} & Phenomenological Parameter \\
\hline
CDW parameters\\
\hline
$s_\beta$ & 0.064 & -- &  Contraction parameter for $\beta$ ring Eq.~\eqref{eq:cntraction} & Fitted to DFT\cite{rossnagel2006spin} \\
$s_\gamma$ & 0.043 & -- &  Contraction parameter for $\gamma$ ring  Eq.~\eqref{eq:cntraction} & Fitted to DFT\cite{rossnagel2006spin}   \\
$\xi$ & $1.4$ & \AA & Wannier orbital width  & Fitted to DFT\cite{rossnagel2006spin} \\
$\Delta$ & $\ve_1 - \ve_0$ & eV & Orbital bias & Phenomenological Parameter   \\
\end{tabular}
\end{ruledtabular}
\end{table*}

\subsection{The Charge-Density Wave \label{ssec:TB-CDW}}

$1T$-TaS$_2$ undergoes a sequence of phase transitions upon lowering the temperature~\cite{wilson1975charge,wang2020correlated}. First, at roughly $T = 360$ K it undergoes an incommensurate CDW transition into a reconstructed metallic state. Subsequently, around $T = 200$ K, a lock-in transition occurs at which the CDW becomes commensurate. This second transition is strongly first order with hysteresis~\cite{wang2020band}. %Moreover, there is a third CDW phase that is observed only upon heating.

\begin{figure}
    \centering
         \includegraphics[width=1\linewidth]{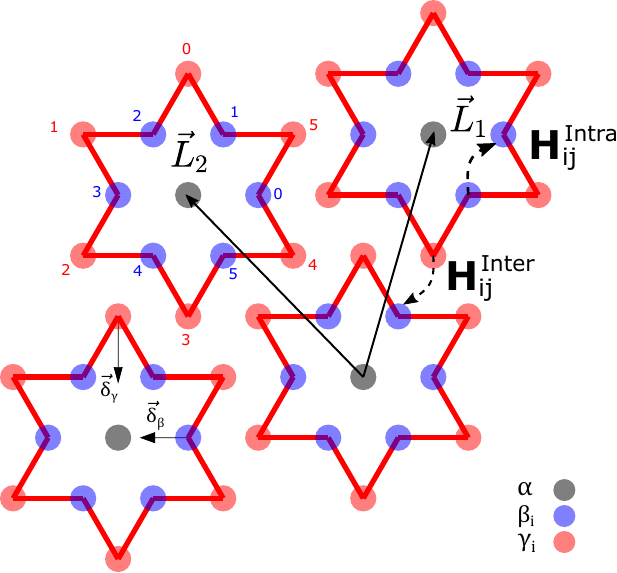}
         \caption{The superlattice structure in the CDW phase. The central atom is denoted by $\alpha$ (grey), the  inner shell is denoted by $\beta_i$ (blue) and the outer shell  is denoted by $\gamma_i$ (red). Each atom on the inner (outer) shell contracts radially with contraction vector $\vec{\delta}_\beta$ ($\vec{\delta}_\gamma$).
        The vectors $\vec{L}_1$ and $\vec{L}_2$ denote the CDW lattice vectors. $H_{ij}^{Intra}$ and $H_{ij}^{Inter}$ denote the hopping matrix elements inside and between CDW sites, respectively. }
         \label{fig:CDWLattice}
\end{figure}

In what follows, we will only be interested in the low temperature commensurate CDW phase. In this commensurate state the crystal is reconstructed and forms a $\sqrt{13}\times\sqrt{13}$ superlattice composed of stars-of-David clusters, as shown in Fig.~\ref{fig:CDWLattice}. Each start-of-David (SoD) consists of 13 Ta atoms, with one central atom and two shells of 6 nearest neighbours (inner shell) and 6 next-to-nearest neighbours (outer shell). These two shells contract radially towards a central Ta atom (see Fig. \ref{fig:CDWLattice}). Consequently, each SoD forms a 13 atoms unit cell in a new triangular superlattice.  The electronic states close to the Fermi energy are dramatically affected by this distortion and exhibit insulating behavior at low temperature, which is believed to be a Mott insulator.

It is important to note that the commensurate CDW phase spontaneously breaks the $\mc C_{2x}$ symmetry, giving rise to two distinct tiling patterns. The sample may be a unidomain or multidomain, in which case domain walls will appear (see STM data presented in Ref.~\cite{Silber2024}). Other types of domain walls are also possible, which correspond to fractional dislocations of the superlattice~\cite{ma2016metallic}.  Recently, voltage driven switching between the two tiling configurations was demonstrated in Ref.~\cite{liu2023electrical}. However, in what follows, we choose a particular tiling pattern and assume the sample is a unidomain and leave the exploration of these intriguing domain walls to future work. The primitive superlattice vectors are chosen as $\vec{L}_1=3\vec{a}_1+4\vec{a}_2$ and $\vec{L}_2=-\vec{a}_1+3\vec{a}_2$. Notice that the axes of the primitive vectors is rotated with respect to the original one $\{\bs a_1,\bs a_2,\bs a_3\}$ by $\approx  13.9\degree$.

To incorporate the effects of the CDW we promote the tight-binding model to a superlattice model by increasing the unit cell, which introduces an additional intra-super cell degree of freedom corresponding to the site number. In accordance with the literature we distinguish three types of sites, denoted $\alpha,\beta,\gamma$ (see Fig.~\ref{fig:CDWLattice}), where $\beta_n$ and $\gamma_n$ with $n=0,\ldots,5$ label the sites in the $\alpha$ and $\beta$ shells. The basis states of the tight tight-binding model are then $|\alpha, d_{0,\pm}\rangle$, $|\beta_n, d_{0,\pm}\rangle$, and $|\gamma_n, d_{0,\pm}\rangle$.  As a result of the radial contraction, the position of the $\beta$ and $\gamma$ sites in the unit cell is modified and takes the form
\be
\Vec{r}_{\beta,\gamma}=\vec{r}_{0,\beta,\gamma}(1-s_{\beta,\gamma})
\ee
where $\Vec{r}_{0}$ is the position before the contraction and the two dimensionless parameters $s_{\beta,\gamma}$ account for the displacement in the CDW state. The values of the displacement are taken from X-ray scattering experiments~\cite{kochubey1998charge} and DFT~\cite{rossnagel2006spin} (see Table \ref{tab:table2}).

We now discuss how the superlattice tight-binding model is lifted from the model for undistorted $1T$-TaS$_2$ introduced above. Three aspects warrant discussion: the on-site couplings (i.e. crystal fields, spin-orbit coupling), the hopping within a cluster, and the hoppings between clusters. We keep the form of the on-site terms the same as before, which is to say that we neglect small deviations within the supercell\footnote{In principle, this potential also depends on the shell $\a,\b$ and $\g$. However, the dependence on shell introduces four additional fitting parameters, which have very little influence on the resulting band-structure and Bloch wave functions. Thus, to prevent over parameterization we omit this dependency.}. Hence, the on site couplings remain $\varepsilon_{0,1}$ and $(\lambda_0,\theta)$. However,  we treat $\D \equiv \ve_1 - \ve_0$ as a parameter and explore the dependence of the miniband electronic structure on the value of the crystal field splitting.

%With the new crystal structure in hand we can now write a tight-binding model. We first write the local (on-site) potential coming from the CDW
%\begin{align}
%     H_{ii}=\begin{pmatrix}
%    \varepsilon_{0} & 0 &0\\
%    0 & \varepsilon_{1} & 0\\
%    0 & 0 & \varepsilon_{1}
%    \end{pmatrix} +H_{SOC}
%    \label{eq:CDWOn}\,.
%\end{align}
%However, it should be noted that the values of $\ve_{0,1}$ may vary due to the formation of the CDW. Moreover, we  wish to explore the properties of the miniband structure away from the precise values obtained in DFT. Therefore, we allow these parameters to vary. In particular, we will consider the orbital bias
%\[ \D \equiv \ve_1 - \ve_0\]
%as a phenomenological parameter.

Next, we turn to the hopping terms, which are modified due to the change in the overlap between neighboring Wannier orbitals of the undistorted lattice.
%We treat each SoD as a single lattice site, by having each site in the SoD as an effective "orbital". In this picture, the only place where the hopping phases come into play is between different SoD sites.
We divide our hopping terms into intra- and inter-unit-cell hopping terms, denoted by $H^{intra}_{ij}$, $H^{inter}_{ij}$. We choose the gauge\footnote{We exploit individual gauge freedom of each site to absorb the phase factor associated with their displacement with respect to the central $\a$ atom.} such that the intra-unit-cell hopping Hamiltonian is k independent and the inter-unit-cell hopping Hamiltonian come with the $k$ dependent phases $e^{i\vec{L}_{ij}\cdot \vec{k}}$, where $\vec{L}_{ij}$ are the lattice vectors connecting lattice sites $i$ and $j$.
Under these conditions, the lowest order correction to the hopping can be written as follows
\begin{align}
    \begin{split}
    &H^{intra}_{ij}=S_{ij}\hat{T}\\
     &   H^{inter}_{ij}(\mathbf{k})=S_{ij}e^{i\vec{L}_{ij}\cdot \vec{k}}\hat{T}
      \end{split}
    \label{eq:CDWHop}
\end{align}
where $\hat T$ is the appropriate hopping matrix along the direction of the $ij$-bond (see Appendix \ref{app:sym} for more details).
% \begin{align*}
%     \hat{T}=\begin{pmatrix}
%     t_1 & -t_{3} &-t_{4}\\
%     -t_{3} & t_0 & -t_{3}\\
%     -t_{4} & -t_{3} & t_1
%     \end{pmatrix}
% \end{align*}
Notice we have introduced a correction to the hopping amplitude encapsulated in the matrix
\be \label{eq:cntraction}
S_{ij}\approx1+ (s_i-s_j) {r}_{ij}^2/\xi^2  \,.
\ee
Here $\xi$ is the  length scale of the Wannier orbital on each site, which we will treat as a phenomenological parameter.

The resulting tight-binding Hamiltonian is a 78$\times$78  matrix (2 spins $\times$ 3 orbitals $\times$ 13 sites) given by
\begin{align}
    H^{CDW} _{ij}(\mathbf{k}) = \left[  H_{ii} \delta_{ij}+ H^{intra}_{i,j}+H^{inter}_{i,j} (\vec{k})\right]\,.
    \label{eq:CDWH}
\end{align}
We note that the  model above includes two phenomenological parameters; the orbitally selective energy shift, $\Delta$, and the extent of the Wannier orbitals, $\xi$. To determine the values of these parameters we fit the resulting flat band and its two nearest neighboring bands to DFT calculations~\cite{rossnagel2006spin}. The results of an unbiased fitting procedure are presented in Fig. \ref{fig:FitDFTBoth}.b and yield $\xi = 1.4\mrm \AA$ and $\D=0.55\,eV$.

\begin{figure*}
    \centering
         \includegraphics[width=1\linewidth]{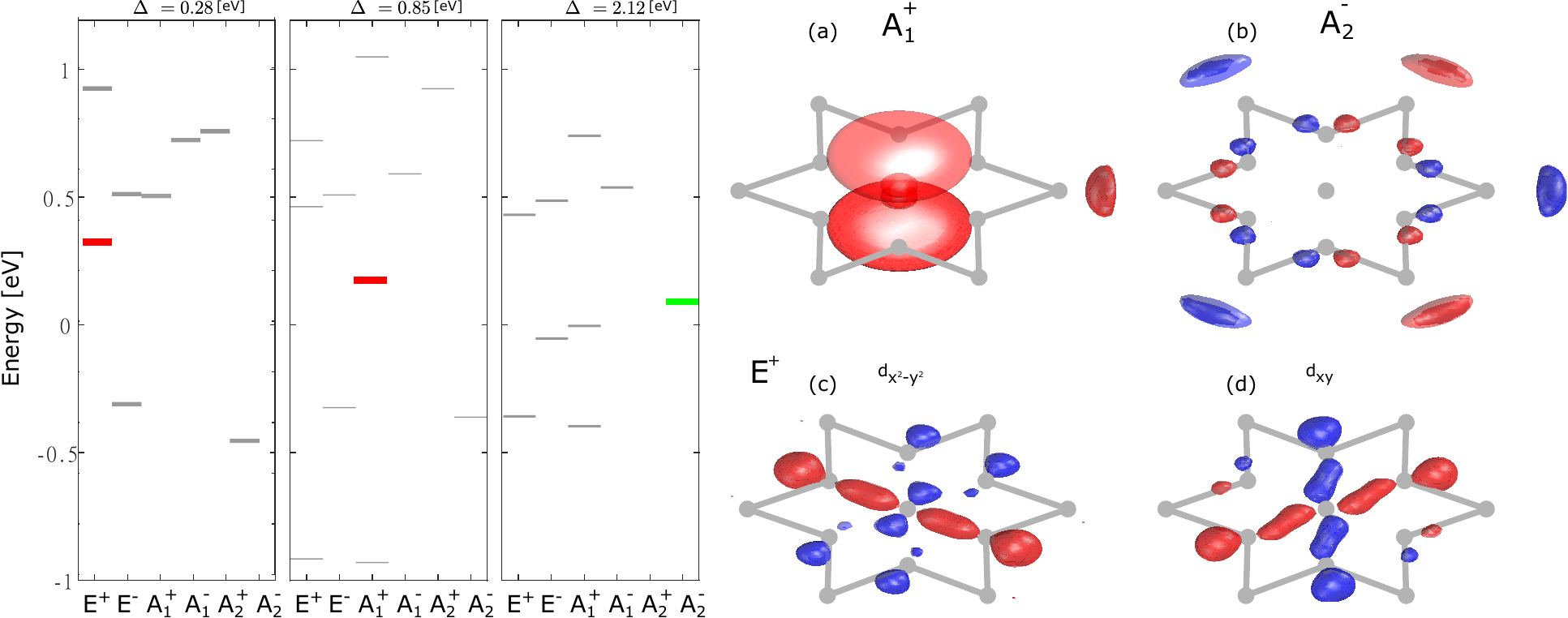}
         \caption{ (a) to (c)
         The energy spectrum of the molecular orbital, for fixed SOC strength $\lambda_0=0$ different values of $\Delta$ [(a) corresponds to $\D = 0.28$ eV, (b) corresponds to $\D = 0.85$ eV and (c) corresponds to $\D = 2.12$ eV]. The state corresponding to seventh band is colored red or green corresponding to whether the cluster state is even or odd under inversion, respectively. (d) to (f) The real space wave-function for each the states indicated in panels (a) to (c) with the corresponding symmetry label. The color  indicates sign of the wave function, and the grey lines represent the SoD cluster.  }

         \label{fig:ClusterStates}
\end{figure*}

\section{The Charge-density wave state viewed as a lattice of weakly coupled clusters}\label{sec:CDW_clusters}

In this section we turn to one of the central results of this paper: an analysis of the flat band electronic structure in terms of weakly coupled SoD clusters. The starting point for this analysis are the isolated clusters (i.e., the basic units of the CDW superlattice), for which we systematically construct symmetrized cluster states. These symmetrized cluster states can be understood as the molecular orbitals of the 13-site units. When expressed in a basis of molecular orbitals, the Hamiltonian of a single SoD cluster is block diagonal and its energy levels can be labeled by the symmetry quantum numbers of the symmetrized states. The symmetry-resolved spectrum of energy levels of a single cluster then allows us to identify the state which forms the flat band, thus providing important insight into the nature of the flat band and its dependence on the parameters of the cluster Hamiltonian.

In constructing the symmetrized cluster states we proceed in two steps. First, we ignore spin-orbit coupling and examine the cluster Hamiltonian in the spin-rotation invariant limit. We then include spin-orbit coupling and form cluster states symmetrized with respect to the double group.

\subsection{Cluster states without SOC \label{ssec:clusters-no-SOC}}

Since a single SoD cluster has the same point group symmetry as the $1T$-TaS$_2$ crystal lattice (in the absence of the CDW), constructing cluster states is a matter of symmetrizing with respect to $D_{3d}$. Without spin-orbit coupling there are two degrees of freedom which must be symmetrized: (i) the sites within a cluster and (ii) the orbitals on each site. The orbitals are symmetrized by default and we therefore first consider the sites within a cluster. We symmetrize these by forming states of definite angular momentum and then construct symmetrized products of site and orbital states.

For the $\beta$ and $\gamma$ sites of the cluster, states of definite angular momentum are given by
\be \label{eq:beta_and_gamma_states}
|\beta,l \rangle  =\sum_{n} \frac{e^{ i \theta nl} }{\sqrt{6}} |\beta_n \rangle, \quad  |\gamma,l \rangle  = \sum_{n} \frac{e^{ i \theta nl}}{\sqrt{6}}|\gamma_n \rangle
\ee
where $l=-2,\ldots,3$ is the angular momentum and $\theta \equiv 2\pi/6$. The central $\alpha$ site coincides with the origin of the cluster and is therefore labeled $l=0$; the corresponding state is denoted $|\alpha, 0\rangle$. The angular momentum states can be assigned symmetry labels (i.e., irreducible representations) according to their transformation properties. Specifically, the $l=0$ states have $A^+_1$ symmetry, the $l=\pm 1$ states have $E^-$ symmetry, and the $l=\pm 2$ states have $E^+$ symmetry. The symmetry labels of the $l=3$ states are different for the $\beta$ and $\gamma$ sites, which is due to different orientation of the six $\beta$ and $\gamma$ sites (see Fig.~\ref{fig:CDWLattice}). It is straightforward to determine that $|\beta, 3\rangle$ has $A^-_1$ symmetry and $|\gamma, 3\rangle$ has $A^-_2$ symmetry.

The next step is to include the orbital degrees of freedom and form symmetrized product states of sites and orbitals. In the language of group theory this can be phrased as decomposing the product representation $(3A^+_1+A^-_1+A^-_2+2E^++2E^-) \times (A^+_1+E^+)$, which is the product of site and orbital representations, into irreducible representations of the group $D_{3d}$. This yields a total of $3\times 13 = 39$ states, of which five have $A^+_1$ symmetry, two have $A^+_2$ symmetry, fourteen have $E^+$ symmetry, three have $A^-_1$ symmetry, three have $A^-_2$ symmetry, and twelve have $E^-$ symmetry. For example, the five $A^+_1$ states are
\begin{align}
|A_1^+ ( 1 ) \rangle &  = |\alpha, 0, d_0 \rangle, \\
|A_1^+ ( 2 )\rangle & = |\beta, 0, d_0 \rangle, \\
|A_1^+ ( 3 ) \rangle &= |\gamma, 0, d_0 \rangle,
\end{align}
which have $l=0$ and only support on $|d_0 \rangle$, and
\begin{align}
|A_1^+ (4) \rangle & = (|\beta, 2, d_+ \rangle+|\beta, \bar 2, d_- \rangle)/\sqrt{2}, \\
|A_1^+ ( 5 ) \rangle & = (|\gamma, 2, d_+ \rangle+|\gamma, \bar 2, d_- \rangle)/\sqrt{2},
\end{align}
which are states constructed such that the total angular momentum is zero. The remaining states of different symmetry are constructed in the same way and are listed in Appendix~\ref{app:cluster}.

As mentioned above, the key upshot of constructing symmetrized cluster states is that such construction block-diagonalizes the cluster Hamiltonian, i.e., the Hamiltonian defined in Sec.~\ref{ssec:TB-CDW} restricted to a single SoD cluster. All energy levels of a single cluster can thus be resolved by point group symmetry. Note that the energies and eigenstates within a given symmetry block are obtained by diagonalization of the block Hamiltonian; due to multiplicities (e.g. the $A^+_1$ states have multiplicity five) symmetrization by point does not fully diagonalize the Hamiltonian. Resolving the cluster states (``molecular states'') by symmetry allows for an analysis of the dependence of the cluster energy levels on model parameters, and therefore provides important insight into the nature of the flat band.

The cluster energy levels obtained by diagonalizing the Hamiltonian within each symmetry block is shown in Fig.~\ref{fig:ClusterStates}(a)--(c) for three different parameter sets. Specifically, we show three different values of the crystal field splitting $\Delta$. Given a filling of one electron per site, the lowest six cluster levels will be fully filled, and the seventh level will be half filled and can thus be interpreted as the flat band state, i.e., the cluster level which will become the half-filled flat band in the CDW state of weakly coupled clusters. Importantly, we see from Fig.~\ref{fig:ClusterStates}(a)--(c) that the nature of the cluster level of interest (indicated in red or green) is different for the three values of the crystal field splitting. In Fig.~\ref{fig:ClusterStates}(a) the relevant cluster level (indicated in red) is a twofold degenerate $E^+$ level, which implies that, in order to obtain a single flat band, spin-orbit coupling is required to split the twofold $E^+$ level. This corresponds to the scenario discussed in Refs.~\onlinecite{rossnagel2006spin} and \onlinecite{he2018spinon}. In contrast, in Fig.~\ref{fig:ClusterStates}(b)  the relevant cluster level (red) is a non-degenerate $A^+_1$ level, in which case a flat band can form in the absence of spin-orbit coupling. This scenario was pointed out in Ref.~\onlinecite{cheng2021understanding}. Upon increasing the value of $\Delta$, we find that a level crossing between the inversion-even $A^+_1$ level and an inversion-odd $A^-_1$ level can occur, suggesting a further scenario for realizing a flat band in $1T$-TaS$_2$ characterized by an even-odd band inversion. This will be explored further in Sec.~\ref{sec:flat-bands}. Although here we only show the dependence of the cluster energy levels on $\Delta$, in general these energy levels also depend on intra- and interorbital hoppings.

%The low energy part of the spectrum of the block diagonal Hamiltonian is plotted in Fig.~\ref{fig:ClusterStates} for three different set of parameters. Panels (a), (b) and (c) correspond to the parameters with $\xi =\sqrt{2} \AA$ and $\Delta = 0.28,\,0.85,\,2.12\, eV$, respectively. For the parameters in panel (a) the 7'th cluster state belongs to the $A_1^+$ irrep, while for panel (b) it belongs to the $E^+$ irrep and $A^-_2$ for panel (c).

To visualize these orbitals in real space we take the weight obtained from the diagonalization process and project them onto real space orbitals of the hydrogen atom in Fig.~\ref{fig:ClusterStates}.(d)-(f), corresponding to the states indicated in panels (a)-(c). Note that for the twofold $E^+$ states here we show real superposition states.

%{\color{red} JV: here we need to change the labeling in figures: a,b,c,etc}

\subsection{Cluster states in the presence of SOC \label{ssec:clusters-SOC}}

When spin-orbit coupling is included the clusters states must be symmetrized with respect to representations of the double group. There are four such representations, which we label $\Gamma^{\pm}_{1/2}$ and $\Gamma^{\pm}_{3/2}$ in accordance with total angular momentum $j=1/2,3/2$ and which all correspond to twofold degenerate representations (due to crystal and time-reversal symmetry).  It is then straightforward to construct spin-oprbit coupled symmetrized states from the states constructed above. All states which belong to non-degenerate representations ($A^\pm_1$ and $A^\pm_2$) fall in the $\Gamma^{\pm}_{1/2}$ symmetry class, which means that some states previously distinct now have the same symmetry. When coupling the $E^\pm$ states with a spin $S=1/2$ both $\Gamma^{\pm}_{1/2}$ and $\Gamma^{\pm}_{3/2}$ states can be formed, since this amounts to coupling an $L=1$ and $S=1/2$ angular momentum to form $j=1/2,3/2$. For instance, taking the $\alpha$ states as an example, we form the states
\begin{align}
|\Gamma^{+}_{3/2},3/2\rangle &= |\alpha,0,d_+,\uparrow\rangle, \\
|\Gamma^{+}_{3/2},\bar 3/2\rangle&= |\alpha,0,d_-,\downarrow\rangle.
\end{align}
Using this standard prescription we find the following multiplicities: $\Gamma^{+}_{1/2}$ (14), $\Gamma^{+}_{3/2}$ (7), $\Gamma^{-}_{1/2}$ (12) and $\Gamma^{-}_{3/2}$ (6). In this basis of symmetrized states the spin-orbit coupled cluster Hamiltonian is block-diagonal.

\begin{figure}
    \centering
    \includegraphics[width=1\linewidth]{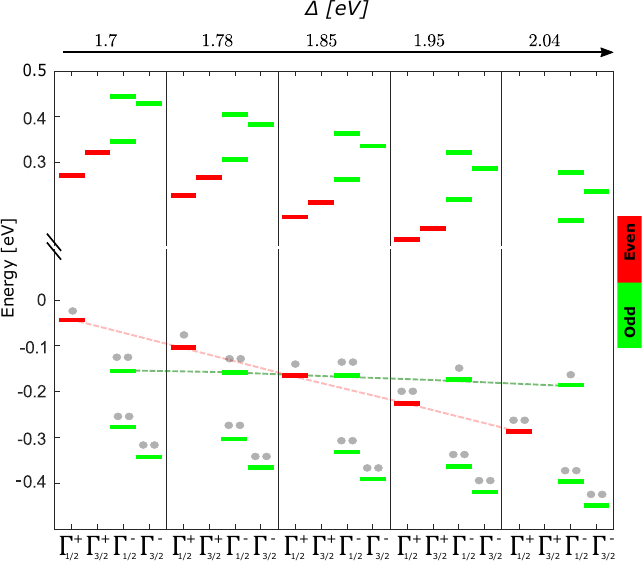}
    \caption{ The spectrum of cluster states for varying values of $\Delta$ for $\t=\pi/4$ [see Eq.~\eqref{eq:H_SO}]. The color of each state denotes its eigenvalue under inversion.  Grey circles indicate level filling.}
    \label{fig:ClusterStatesEnergy}
\end{figure}

In Fig.~\ref{fig:ClusterStatesEnergy} we plot the energy spectrum of the spin-orbit coupled cluster states as a function of the orbital detuning $\D$ for $\t = \pi/4$. The grey dots denote the electronic ground state occupation in the absence of interaction, which corresponds to 13 filled electronic states. As a result, there is one singly occupied state, which potentially forms the flat band when the SoDs form a lattice structure.

The different panels in the figure depict the  spectral flow of these states as a function of the parameter $\D$. The even- and odd-parity states are labeled red and green, respectively.
As can be seen, the spectral flow also involves an inversion in between even and odd parity states.  In the regime
between  $\D < 1.78$ the 7th energy state is the even parity state $\G_{1/2}^+$. Above this threshold it switches to the odd parity $\G_{1/2}^-$.
As we will see in the next section such a transition lead to transitions between even and odd flat bands with an intermediate topologicallly non-trivial  band.

% It is important to note that the CDW breaks the $\mc C_{2x}$ symmetry, but does not break mirror-$z$ and inversion.
\begin{figure*}[t]
    \centering
    \includegraphics[width=0.9\linewidth]{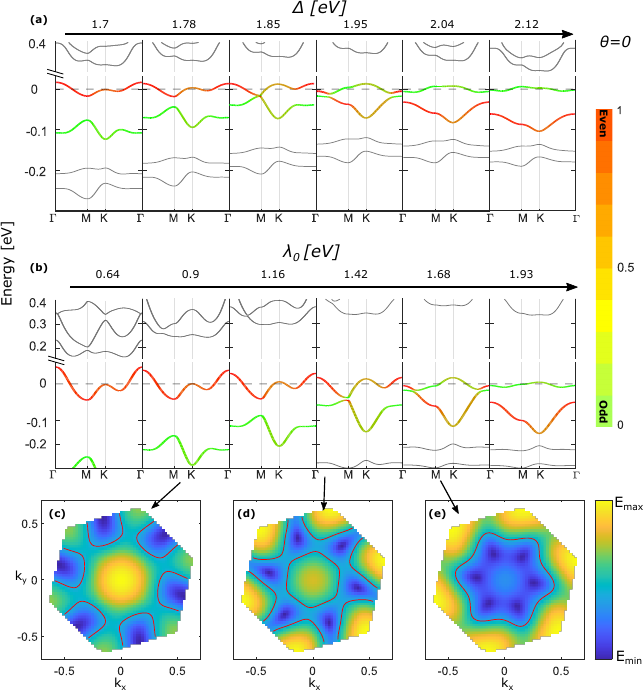}
    \caption{ (a) The dispersion in the mini-Brillouin zone for varying values of $\D$, $\lambda_0 = 0.64$ eV and $\t = 0$. (b) Same for varying $\lambda_0$ and $\D = 0.13$ eV. The color indicates the relative weight of even and odd orbitals in the Bloch wave function.
    (c) to (e) are color maps of the dispersion of the flat band at indicated values of $\lambda_0$. The red lines indicate the Fermi surface at half-filling. }
    \label{fig:FlatBandEnergy}
\end{figure*}

\section{Flat band inversion \label{sec:flat-bands}}

The tight binding model we have constructed contains two phenomenological parameters, the orbital splitting $\Delta$ and the Wannier orbital spread $\xi$. To determine realistic values we used an unbiased fit to DFT. In addition to these parameters, we have also introduced the SOC parameters $\lambda_0$ and $\theta$, which control the overall and relative strength of diagonal and off-diagonal elements of the SOC matrix, respectively. To gain further insight on the nature of the flat band in 1$T$-TaS$_2$, we will now explore the influence of these parameters on the miniband structure by allowing them to vary. (We generally find that  the parameter $\xi$ is inconsequential over the range of physically relevant parameters and therefore mainly our focus to  the parameters $\Delta$, $\t$ and $\lambda_0$.)

As mentioned in the previous section Sec.~\ref{sec:CDW_clusters}, the flat miniband of interest stems from the seventh cluster state, which is half-filled. We have furthermore demonstrated that the energy levels of the cluster states (when considering single clusters) depends on the orbital bias $\D$, and also on the SOC parameters $\lambda_0$ and $\t$. As a result, the structure of the energy level spectrum, and in particular the nature of the flat band state, can vary depending on these three parameters. It also depends on the hopping overlap integrals, but here we focus attention on the dependence on $\D$, $\lambda_0$, and $\t$.

When $\D=0$ and $\lambda_0=0$, the seventh cluster state belongs to a fourfold degenerate energy level (including spin), which has $E^+$ symmetry under $D_{3d}$. Finite SOC will then split this fourfold level into two twofold energy levels with $\Gamma_{1/2}^+$ and $\Gamma_{3/2}^+$ symmetry, leading to half-filled ``band''. Hence, in this scenario, a flat band arises by allowing hopping between the $\Gamma_{1/2}^+$ or $\Gamma_{3/2}^+$ states on neighboring clusters. As alluded to in Sec.~\ref{sec:CDW_clusters}, this is one way to obtain the half-filled flat band in the miniband structure consistent with experiment. In this scenario SOC plays a crucial role. Indeed, this scenario is the one proposed by Ref.~\cite{rossnagel2006spin} and corresponds to the minimal model considered in Ref.~\cite{he2018spinon}.

However, we find additional paths to obtaining a single flat band that do not require SOC.  For example, by tuning the orbital potential $\D$ at $\lambda_0 = 0$ above a certain threshold, the seventh cluster state switches its symmetry between the $E^+$ to $A_{1}^+$ irreps. Upon further increasing $\D$ it switches again to the $A_2^-$ irrep (see Fig.~\ref{fig:ClusterStates}). Given that the two latter states belong to a one-dimensional irrep, they provide a different path to obtain a isolated flat band that does not require SOC. Indeed, we find such flat bands in our tight-binding model, even when the SOC coupling is set to zero.

Guided by this microscopic picture we now explore the miniband structure  as we tune $\D$, $\lambda_0$ and $\t$. We classify three distinct flat bands. Two of these bands, are topologically trivial and classified by the parity of the wave functions at both
high symmetry points, $\Gamma$ and $M$. The third type of flat band is a topologically non-trivial band with opposite parity at the $\Gamma$ and $M$ points.

\subsection{The Ising limit ($\theta=0$)}

The limit of $\theta=0$, although unphysical, is a good starting point to gain intuition. In this limit, the SOC conserves the spin projection along the $z$ axis ($S_z$) and the tight-binding Hamiltonian \eqref{eq:CDWH} becomes  block-diagonal, where the two  blocks of size 39$\times$39 are identical except for an energy shift  due to SOC Eq.~\eqref{eq:H_SO}.

% As mentioned, to analyze the bands we use cluster states (see  Appendix \ref{app:sec:ClusterOrbits}).
% cluster state. The orbital content is predominantly  $d_{\pm}$, which is consistent with Refs.~\cite{rossnagel2006spin,he2018spinon}. The spatial distribution is predominantly the  central $\alpha$ atom, such that most of the weight of the Wannier orbitals is localized close to the center of the SoD.
% The orbital content of the band can be tuned by the orbitally selective potential $\D_\epsilon$ and the spin-orbit coupling strength $L_{SOC}$.
% predominantly  $d_0$. Thus, the corresponding cluster state belongs to a different irrep, $A_1^+$.
% {\color{red} [Amir: The discussion here is very qualitative. Why not have some figures of the orbital content? Also, is this state mostly on $\alpha$ as well?]}
% As  the value of the potential  is increased further, above $\Delta_?>?$, a third type of flat band emerges. As before the $\Gamma$-point wavefunction is predominantly $d_0$. However, its spatial structure belongs to the $A_2^-$ representation of cluster states. This band is thus distinct from the two bands mentioned before. Namely, it is \textit{odd} under inversion at the $\Gamma$-point. Consequently, its weight on the  central $\alpha$ atom vanishes, and thus the Wannier orbital of this type of flat band is extended compared to the previous examples.

In Fig.~\ref{fig:FlatBandEnergy} we plot the flat band's dispersion along the trajectory $\Gamma$--$M$--$K$--$\Gamma$, for different values of $\Delta$ (panel a) and $\lambda_0$ (panel b).
The color coding indicates the weight of even/odd cluster states in the Bloch wave-function. Namely, green indicates the weight of odd parity cluster states and red indicates the weight of the even parity ones.

The parity eigenvalues at the high symmetry points, $\Gamma$ and $M$ can be used to classify the bands into three groups: (i) When the parity at $\G$ and $M$ is even the band will be referred to as ``even'' and the Wannier orbital is even under inversion. (ii) When both $\G$ and $M$ are odd the band will be referred to as ``odd'' and the Wannier orbital is odd under inversion. (iii) When the parity at $\G$ and $M$ is opposite the band has a non-trivial $\mathbb Z_2$ topological invariant~\cite{kane2005z,fu2007topological}. In this case a time-reversal invariant  Wannier representation of the band is obstructed~\cite{soluyanov2011wannier} and thus the orbitals are expected to be extended.

As can be seen in Fig.~\ref{fig:FlatBandEnergy}(a) tuning the potential $\D$ drives a transition between even and odd bands with an intermediate topologically non-trivial band. The cluster state classification of the $\G$-point wave function in this case is either $A_1^+$ or $E^+$ depending on parameters (See Fig.\ref{fig:ClusterStates}).
On the other hand, the odd band  belongs to the $A_2^-$ irrep.

The same behavior is observed upon increasing the spin-orbit coupling strength $\lambda_0$ at constant $\D$ (panel b).
In panels (c)-(e) we plot the band dispersion (and the Fermi surface at half-filling) for three indicated values. Mirror symmetry breaking of the SoD CDW is manifest in the dispersion relations. Namely, they have a ``swirling'' structure reminiscent of a Plumeria flower.

\subsection{The isotropic case ($\theta= \pi/4$)}
When $\theta\neq0$, the spin projection along $z$ is no longer a good quantum number.  In what follows we consider the isotropic limit where $ \t = \pi/4$, which corresponds to the limit where the coordination of sulfur atoms surrounding the tantalum is a perfect octahedron and the $d$-orbitals form an effective $l=1$ irrep.

\begin{figure}
    \centering
    \includegraphics[width=0.9\linewidth]{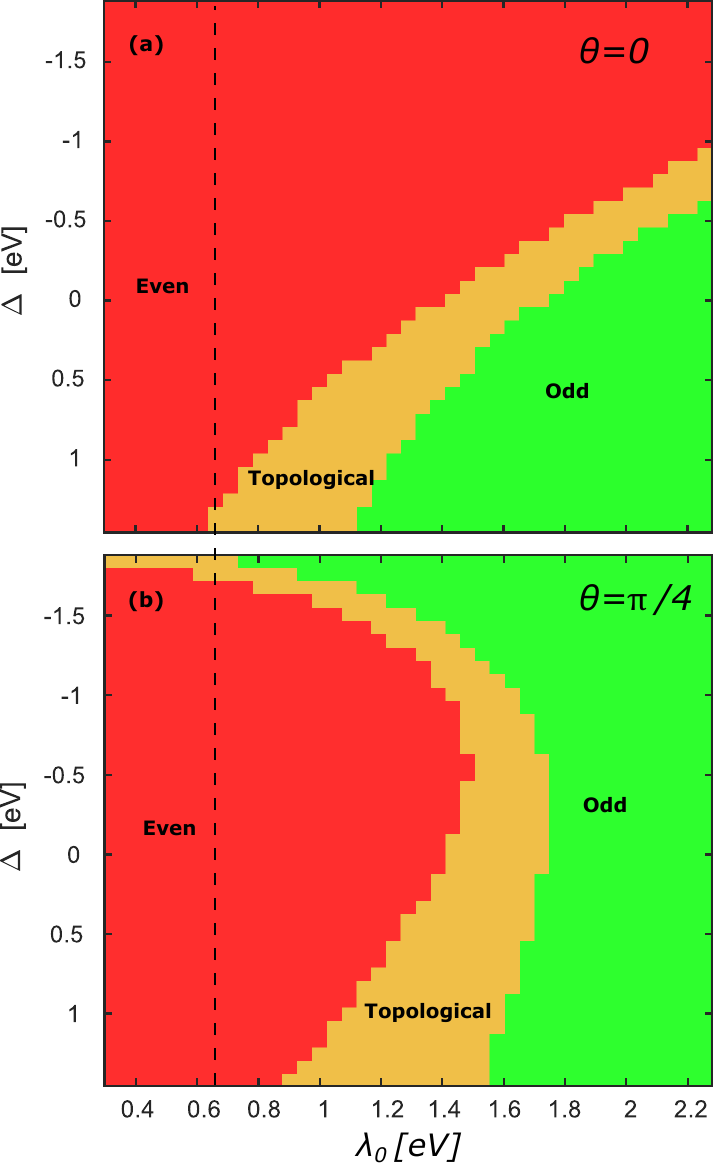}
    \caption{The phase diagrams of the flat band as a function of $\Delta$, and SOC strength, $\lambda_0$.  Red and green indicate bands which are even and odd at both high symmetry points, respectively (see text). The orange area denotes regions where the band has a non-trivial topological index. (a) The phase diagram for $\t = 0$ and (b) for $\t = \pi/4$.}
    \label{fig:Topological}
\end{figure}

In Fig.~\ref{fig:Topological} we plot the classification of the  flat band as a function of $\lambda_0$ and $\D$. We compare the phase diagram of the Ising case ($\t=0$) case (panel a) with the isotropic case ($\t = \pi/4)$ (panel b). As can be seen, the phase space of the topological band is enhances in the isotropic case. Secondly, the odd parity band is promoted to the  $\D>0$ region as well.

% \subsection{Topologically non-trivial flat band }
As we have mentioned above the existence of inversion symmetry endows the band with a $\mathbb Z_2$ topological index. When the Bloch wave-fucntion has different inversion eigenvalues at the high symmetry points $\G$ and $M$ the index is $1$ and the band is topologically non-trivial~\cite{kane2005z}.

%This existence of a topological regime in the phase diagram is of special interest for multiple reasons. First, when the topological index is non-trivial there is no local Wannier representation without breaking time-reversal symmetry. Therefore, the fate of the strongly correlated band at half filling is not clear. Second, in this case the band  potentially has  Berry curvature. Finally, there will be topological edge states running between the flat band and the  bands below or above it. The two latter points are especially relevant to experiments in 4Hb-TaS2 and 1H/1T bilayers, where charge transfer between the layers has been observed, both experimentally and numerically~\cite{nayak2021evidence,ShiweiShen:77401,crippa2023heterogeneous,nayak2023first}. This results in a doped flat band, which might realize  metallic and or even band-insulating behavior.

\section{Discussion \label{sec:discuss}}

We have derived a tight-binding model for 1T-TaS$_2$, which includes the effects of spin-orbit coupling and the star-of-David charge density wave. We found that a single half-filled flat band at the Fermi level is a robust feature of this model in the sense that it exists over a wide range of parameters.
The flat band can be obtained with or without spin-orbit coupling
and may differ in its symmetry properties and even its topological classification.
These results affect the low-energy Hamiltonian in the half-filled Mott state in pristine 1T-TaS$_2$ and also to doped samples, such as 4Hb-TaS$_2$~\cite{nayak2021evidence} and ionic gel gating~\cite{yu2015gate}.

First, the off-diagonal elements in the SOC matrix included in our tight-binding model Eq.~\eqref{eq:H_SO} entail the absence spin  conservation.  The authors of Ref.~\onlinecite{he2018spinon} derived an effective low-energy spin-spin Hamiltonian in the Mott state while  assuming this symmetry is present. Consequently, they arrived at an $XXZ$ Hamiltonian (up to nearest neighbour couplings)
\begin{align}
H_{XXZ} = \sum_{\langle ij\rangle}\left[J_\perp\left(S_i^x S_j^x+S_i^y S_j^y\right)+ J_{ZZ} S^z_i S^z_j \right]\,.
\end{align}
However, once spin conservation symmetry is absent a more general Hamiltonian is allowed~\cite{Maksimov2019}
\begin{align}\label{eq:J-K-Gamma}
\begin{split}
H = H_{XXZ}&+\sum_{j}\sum_{l=1}^3\bigg[ J_1\left( \cos\phi_l \mathcal T_{j,\bs a_l}^{xz}+\sin \phi_l  \mathcal T_{j,\bs a_l}^{yz}\right)\\&+J_2\left( \cos2\phi_l (\mathcal T_{j,\bs a_l}^{xx}-\mathcal T_{j,\bs a_l}^{yy})+\sin 2\phi_l  \mathcal T_{j,\bs a_l}^{xy}\right)\bigg]\,,
\end{split}
\end{align}
where $\phi_l = \phi_0 + 2\pi l/3$ and $\mathcal T_{j,\bs a_l}^{\alpha \beta} \equiv (S_j^\alpha S_{j+\bs a_l}^\beta + S_{j}^\beta S_{j+\bs a_l}^\alpha)/2$.

The Hamiltonian Eq.~\eqref{eq:J-K-Gamma} is a generalized version of the Heisenberg-Kitaev~\cite{jackeli2009mott} model (also known as the $J$--$K$--$\Gamma$--$\Gamma'$ model) on a triangular lattice~\cite{Maksimov2019,wang2021comprehensive}.
The values of the couplings $J_1$, $J_2$ and $\phi_0$ can not be determined based on symmetry alone and require a microscopic calculation~\cite{kruger2009spin}.
It is important to note that a finite angle $\phi_0$ is unique to 1T-TaS$_2$. It reflects the breaking of $\mathcal C_{2x}$ by the SoD CDW. The ground state of this model depends strongly on the parameters and may include spin-liquid ground states~\cite{Maksimov2019}.

As we have seen the topological classification of the flat band can be tuned using the parameters $\D$, $\lambda_0$ and $\t$. The existence of a topologically non-trivial flat band in the phase space is interesting for multiple reasons. First, when the topological index is non-trivial there is no local Wannier representation without breaking time-reversal symmetry. Therefore, the fate of the strongly correlated band at half filling is not clear~\cite{Bernevig2006,Neupert2011,neupert2015fractional,peri2021fragile,turner2022gapping,wagner2023mott}.
Second, there will be topological edge states running between the flat and remote bands, which may be observed even in the presence of bulk states.
Finally, quantum geometrical effects are expected to be strong even in the metallic states.  The latter point is especially relevant to samples where the density is doped away from half-filling. This can be obtained by ionic liquid gel gating~\cite{yu2015gate} or naturally occur due to work function differences to adjacent two-dimensional layers, as observed in 4Hb-TaS$_2$~\cite{nayak2021evidence,ShiweiShen:77401,crippa2023heterogeneous,nayak2023first}. It is thus interesting to consider the Fermi surface instabilities of the doped flat miniband in 1T-TaS$_2$ (see panels (c)-(e) of Fig.~\ref{fig:FlatBandEnergy}). In this case we may expect spontaneous symmetry breaking states such as ferromagnetism or electronic ferroelectricity~\cite{sodemann2017quantum,feldman2016observation}.

\section{Acknowledgments}
We thank Roser Valenti, Binghai Yan, Natalia Perkins, Assa Aurbach, Amit Kanigel and Anna Keselman for helpful discussions. J.R. was funded by the Israeli Science Foundation grant No. 3467/21. J.W.F.V. was supported by the National Science Foundation Award No. DMR-2144352.
%\bibliography{TMDs}
% -> Premade bib

%
% <- Premade bib

\appendix
% \onecolumngrid

\section{Crystallographic and octahedral coordinate systems \label{app:basis}}

\begin{figure}
    \centering
         \includegraphics[width=\columnwidth]{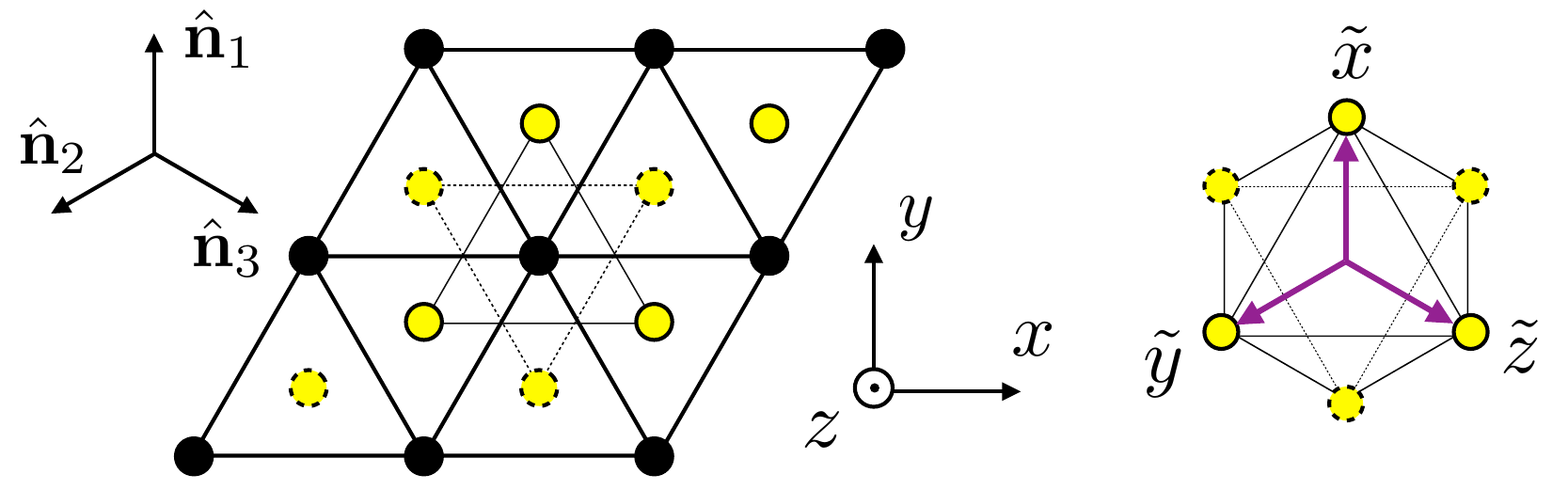}
         \caption{Relationship between coordinate systems used in this paper. Left: The crystal axes from above $x,y,\,\&z$ relative to the crystal structure. Black dots are Ta atoms while yellow ones are S atoms. The solid vs. dashed border distinguishes between S atoms above and below the plane, respectively. Right: The octahedral coordinate system $\tilde x,\tilde y,\,\&\tilde z$. Notice that all three have a positive projection on the $z$ axis. }
         \label{fig:coordinates}
\end{figure}

In this Appendix we present details of the mapping between the crystallographic and octahedral coordinate systems. As explained in the main text, the crystallographic coordinate system is anchored to the crystallographic plane of the triangular Ta atoms, whereas the octahedral coordinate system is anchored to the octahedrally coordinated S atoms surrounding a Ta atom. This is shown schematically in Fig.~\ref{fig:coordinates} (left panel).

We first define the rotation matrix which relates the two coordinate frames. We define three vectors $\hat \bn_{j=1,2,3}$ as
\be
\hat \bn_j =  -\sin\theta_j \hat\bx +  \cos\theta_j \hat\by, \quad \theta_j = (j-1)\frac{2\pi}{3},
\ee
which are shown in Fig.~\ref{fig:coordinates} and are useful to define the unit vectors of the octahedral frame, also shown in Fig.~\ref{fig:coordinates} (right panel). Note that for illustration purposes here we have assumed that the S atoms surrounding the Ta atoms form a perfect octahedron. The unit vectors are then given by
\be
\ehat_{\tilde x, \tilde y, \tilde z} = \sqrt{\frac{2}{3}}\left(\hat \bn_{1,2,3}+ \frac{1}{\sqrt{2}}\hat\bz\right),
\ee
and this triad of unit vectors can be collected in the rotation matrix $R$ as
\be
R = \begin{pmatrix}\ehat_{\tilde x} &\ehat_{\tilde y} & \ehat_{\tilde z} \end{pmatrix}.
\ee
Now $R$ defines a basis transformation relating the crystallographic and octahedral coordinates. In particular, any vector $\br$ can be expanded as
\be
\br = x \hat\bx + y \hat\by + z\hat\bz = \tilde x \ehat_{\tilde x}+\tilde y \ehat_{\tilde y}+\tilde z \ehat_{\tilde z},
\ee
from which it follows that the coordinates are related as
\be
\begin{pmatrix}x\\y\\z \end{pmatrix} = R \begin{pmatrix}\tilde x\\ \tilde y\\ \tilde z \end{pmatrix} . \label{app:R-mat}
\ee

Having discussed how the coordinates in the two frames are related, we next consider the relationship between the crystallographic and octahedral basis for the three relevant $d$-orbitals. In the crystallographic frame---which is the basis used in the main text---the three $d$ states are denoted $(| d_0 \rangle ,| d_\pm \rangle )$ and in the octahedral frame the $d$ states are denoted $(| d_{\tilde y \tilde z}\rangle,| d_{\tilde z \tilde x}\rangle,| d_{\tilde x \tilde y}\rangle)$ and will be referred to as the octahedral $t_{2g}$ states. The state $| d_0 \rangle$ is then expressed in terms of the $t_{2g}$ states as
\be
| d_0 \rangle = \frac{1}{\sqrt{3}}(| d_{\tilde y \tilde z}\rangle +| d_{\tilde z \tilde x}\rangle +| d_{\tilde x \tilde y}\rangle),
\ee
and the two states $|d_\pm \rangle$ are given by
\be
|d_\pm \rangle = \frac{-i}{\sqrt{3}}(|d_{\tilde y \tilde z}\rangle+e^{\pm 2\pi i /3}| d_{\tilde z \tilde x}\rangle+e^{\mp 2\pi i /3}|d_{\tilde x \tilde y}\rangle).
\ee
These relations can be collected in a basis transformation matrix $U$ given by
\be
U = \frac{1}{\sqrt{3}} \begin{pmatrix} -i  & 1 & -i  \\-i\omega  & 1 & -i\omega^2 \\-i\omega^2  & 1 & -i\omega \end{pmatrix}. \label{app:U-mat}
\ee
One way to obtain this basis transformation matrix is to start from a set of three $t_{2g}$ orbitals $(\tilde y \tilde z,\tilde z \tilde x,\tilde x \tilde y)$ expressed in octehedral coordinates, and re-express them in terms of the crystallographic coordinates $(x,y,z)$ using \eqref{app:R-mat}. A subsequent basis transformation $U$ in the space of orbitals then yields the crystallographic orbitals $2z^2-x^2-y^2$ (corresponding to the state $| d_0 \rangle$) and $i(x\mp i y)^2 \pm \sqrt{2}z(x\pm iy)$ (corresponding to the states $| d_\pm \rangle$). It is important to stress that this assumes a perfect octahedral environment of the Ta atoms. In the case of trigonally distorted TaS$_6$ octahedra the $| d_\pm \rangle$ states are some linear combination of the orbitals $ i(x\mp i y)^2 $ and $\pm z(x\pm iy)$, which have the same symmetry under $D_{3d}$ and can therefore mix with coefficients not fixed by symmetry.

We note that this is rather different for $1H$ TMDs with $D_{3h}$ symmetry. The orbitals $ i(x\mp i y)^2 $ and $\pm z(x\pm iy)$ have distinct symmetry under $D_{3h}$ symmetry and therefore cannot mix. This has consequences for the form of spin-orbital coupling, for instance, as pointed out in the main text and further discussed in Appendix~\ref{app:soc} below.

\section{Crystal lattice symmetries \label{app:sym}}

The symmetries of the crystal lattice play a key role in the analysis presented in the main text. This Appendix discusses further details of the crystal symmetry properties of $1T$-TaS$_2$, in particular the representations of the crystal point group in the space of the three $d$-orbitals.

As mentioned in the main text, pristine $1T$-TaS$_2$ has trigonal point group $D_{3d}$, which is generated by the three-fold rotation $\mathcal{C}_{3z}$, two-fold rotation $\mathcal{C}_{2x}$, and the inversion $\mathcal{I}$. (Note that the two-fold rotation $\mathcal{C}_{2x}$ is broken in the commensurate CDW state.) For reference, the character table for the group $D_{3d}$ is reproduced in Table~\ref{tab:CharTable}.
All three states $|d_0 \rangle $ and $|d_\pm \rangle $ are invariant under inversion. The state $|d_0 \rangle $ is furthermore invariant under the threefold and twofold rotations $\mathcal{C}_{3z}$ and $\mathcal{C}_{2x}$, and therefore transforms as $A_1^+$. The states $|d_\pm \rangle $ transform under rotations as
\be
\mathcal{C}_{3z} : |d_\pm \rangle \rightarrow e^{\mp 2\pi i/3}|d_\pm \rangle, \quad  \mathcal{C}_{2x} : |d_\pm \rangle \rightarrow |d_\mp \rangle, \label{app:symmetries}
\ee
and thus transform as $E^+$. As a result, the matrix representations of the rotation symmetries in the space of orbitals are given by
\be
\mathcal{C}_{3z} = \begin{pmatrix} \omega^2  & 0 & 0  \\ 0  & 1 & 0\\  0  & 0 & \omega \end{pmatrix}, \quad  \mathcal{C}_{2x} = \begin{pmatrix} 0  & 0 & 1  \\ 0  & 1 & 0\\  1  & 0 & 0 \end{pmatrix} .\label{app:sym-matrix}
\ee
Note that the matrix representation of the inversion $\mathcal{I}$ is equal to the identity. In addition $D_{3d}$ point group symmetry, both the normal state and the CDW state of $1T$-TaS$_2$ have time-reversal symmetry $\mathcal T$. Time-reversal leaves $|d_0 \rangle $ invariant but acts on $|d_\pm \rangle$ as
\be
\mathcal T : |d_\pm \rangle \rightarrow -|d_\mp \rangle. \label{app:T-sym}
\ee

\begin{table}
\caption{\label{tab:CharTable}
Character table of $D_{3d}$ point group}
\begin{ruledtabular}
\begin{tabular}{cccccccc}
 $D_{3d}$ &$\mathbb{1}$& $\mathcal{C}_{3z}$ & $\mathcal{C}_{2x}$ & $\mathcal{I}$ &  $\mathcal{I}\mathcal{C}_{3z}$ & $\mathcal{M}_{2x}$ \\
\hline
$A^{\pm}_1$ & 1 & 1 & 1 & $\pm1$ & $\pm1$ & $\pm1$\\
$A^{\pm}_2$ & 1 & 1 & $-1$ & $\pm1$&   $\pm1$ & $\mp 1$\\
$E^{\pm}$   & 2 & $-1$ & 0 & $\pm2$  & $\mp 1$ & $0$
\end{tabular}
\end{ruledtabular}
\end{table}

These symmetries can be used to determine the form of Hamiltonian in Eq.~\eqref{eq:TrigH}, in particular the allowed hoppings. Consider the nearest neighbor hopping matrix in the direction of $\mathbf{a}_1$ (see Fig.~\ref{fig:CrystalStructure}), which we denote $T_1$. Inversion symmetry and the twofold rotation $\mathcal C_{2x}$ impose the constraints $T_1 = T^\dagger_1$ and $\mathcal{C}_{2x}  T_1 \mathcal{C}^\dagger_{2x} =T_1$, respectively. Time-reversal symmetry gives an additional constraint, leading to the most general form of the hopping matrix given by
\be
T_1 = \begin{pmatrix} t_2  & i t_3 & t_4  \\ - i t_3  & t_1 & -i t_3 \\  t_4  &  i t_3  & t_2 \end{pmatrix}.
\ee
Here $t_{1,2,3,4}$ are the four independent real hoppings parameters used to parametrize the Hamiltonian Eq.~\eqref{eq:TrigH}. The hopping matrices $T_{2,3}$ corresponding to the bond directions $\mathbf{a}_{2,3}$ are obtained by applying the threefold rotation $\mathcal C_{3z}$.

It is straightforward to determine the form of the symmetry operators in the octahedral basis. Making use of the basis transformation matrix $U$ given by \eqref{app:U-mat}, we find
\begin{align}
\widetilde{\mathcal{C}}_{3z} =U \mathcal{C}_{3z} U^\dagger & =\begin{pmatrix} 0 & 1 & 0  \\ 0  & 0 & 1\\  1  & 0 & 0 \end{pmatrix}, \\
\widetilde{\mathcal{C}}_{2x} =U \mathcal{C}_{2x} U^\dagger &= \begin{pmatrix} 1  & 0 & 0  \\ 0  & 0 & 1 \\  0  & 1 & 0 \end{pmatrix},
\end{align}
which is consistent with what would be expected for three $t_{2g}$ orbitals with $(\tilde y \tilde z, \tilde z \tilde x, \tilde x \tilde y)$ symmetry.

The hopping matrix $T_1$ may also be expressed in terms of the octahedral basis. Symmetry considerations dictate that the hoppings matrix in the octahedral basis, denoted $\widetilde T_1 $, takes the form
\be
\widetilde T_1  = \begin{pmatrix} t_a  & t_c & t_c  \\ t_c  & t_b & t_d \\  t_c  &  t_d  & t_b \end{pmatrix},
\ee
where $t_{a,b,c,d}$ are four hopping parameters in the octahedral basis. Setting this expression for $\widetilde T_1 $ equal to $U T_1 U^\dagger$, we obtain a relationship between the two different hopping parametrizations. Specifically, we find
\be
\begin{pmatrix}t_a \\ t_d \\ t_c \\ t_d \end{pmatrix} = \frac13 \begin{pmatrix} 1  & 2 & 4 & 2  \\ 1  & 2 & -2 & -1  \\ 1  & -1 & 1 & -1 \\ 1  & -1 & -2 & 2  \end{pmatrix} \begin{pmatrix}t_1\\ t_2 \\ t_3 \\ t_4 \end{pmatrix}.
\ee
Note that the matrix which relates the two sets of hopping parameters is its own inverse.

\section{Spin-orbit coupling in $D_{3d}$ and $D_{3h}$}
\label{app:soc}

This Appendix elaborates on the discussion of spin-orbit coupling in TMD materials (Sec.~\ref{ssec:TB-SOC}), in particular the differences between the $D_{3d}$ and $D_{3h}$ structures.

The explicit form of the orbital angular momentum matrices $\bL = (L_x,L_y,L_z)$ used in Eq.~\eqref{eq:H_SO} is given by
\be
L_z = \begin{pmatrix} 1 & 0 & 0  \\ 0  & 0 & 0\\  0  & 0 &-1\end{pmatrix}, \quad  L_+ = L^\dagger_- = \begin{pmatrix} 0  & \sqrt{2} & 0  \\ 0  & 0 & \sqrt{2} \\  0  & 0 & 0 \end{pmatrix} \label{app:L-mat}
\ee
where we used the basis $(| d_+ \rangle, | d_0 \rangle,| d_- \rangle )$ for the orbitals. Note that in this basis the orbital angular momentum matrices take the standard form of an $S=1$ spin angular momentum and have the expected transformation properties under all symmetries. As mentioned in the main text, the parameter $\theta$ introduced in Eq.~\eqref{eq:H_SO} is related to the orbital content of the three states valence and conduction states $(| d_+ \rangle, | d_0 \rangle,| d_- \rangle )$. In particular, in the Ising limit ($\theta=0$) these three states exactly correspond to the orbitals $i (x \mp i y)^2$ ($d_\pm$) and $2z^2-x^2-y^2$ ($d_0$), here expressed in crystallographic coordinates.

The Ising limit is therefore applicable to the $1H$ structures with point group $D_{3h}$, since $D_{3h}$ symmetry forbids mixing between the $i (x \mp i y)^2$ orbitals and the $\pm z(x\pm iy)$ orbitals (see also Appendix~\ref{app:basis}). Indeed, the $i (x \mp i y)^2$ orbitals are even under the horizontal mirror reflection $\sigma_h$, whereas the $\pm z(x\pm iy)$ orbitals are odd. As was noted previously in Ref.~\onlinecite{liu2013three-band}, when projecting the isotropic atomic spin-orbit coupling Hamiltonian of the full five $L=2$ $d$-states into the subspace of the three orbitals given by $d_{x^2-y^2}$, $d_{xy}$, and $d_{z^2}$, only the Ising term survives.

In the $1T$ structures with point group $D_{3d}$ the local environment of the Ta atoms is very different. Indeed, rather than a triangular prism, as in the $1H$ structures, the S atoms form a (distorted) octahedron. This implies that in general the $i (x \mp i y)^2$ and $\pm z(x\pm iy)$ orbitals will mix, since they have the same symmetry under $D_{3d}$. Now, the spin-flip terms in $H_{\text{SO}}$ need not vanish, giving rise to the more general form of Eq.~\eqref{eq:H_SO}. Since the degree of mixing is material-specific, so is the parameter $\theta$.

The orbital angular momentum matrices can also be expressed in the octahedral basis. Denoting the $\tilde z$ component as $\widetilde L_{\tilde z} $, one has
\be
\widetilde L_{\tilde z} = U \ehat_{\tilde z}  \cdot \bL U^\dagger,
\ee
where $U$ was defined in Eq.~\eqref{app:U-mat}. Similar expressions hold for the $\tilde x$ and $\tilde y$ components. In the octahedral basis the orbital angular momentum matrices have the familiar form
\be
\widetilde L_{\tilde z} = \begin{pmatrix} 0 & -i & 0  \\ i  & 0 & 0\\  0  & 0 &0\end{pmatrix}, \widetilde L_{\tilde x} = \begin{pmatrix} 0 & 0 & 0  \\ 0  & 0 & -i \\  0  & i & 0 \end{pmatrix},  \widetilde L_{\tilde y} = \begin{pmatrix} 0 & 0 & i  \\ 0  & 0 & 0 \\  -i  & 0 & 0 \end{pmatrix},
\ee
which emphasizes the equivalence of the octahedral axes.

\section{Construction of cluster states \label{app:cluster}}

% \be
% |\beta,l \rangle= \frac{1}{\sqrt{6}}\sum_{n}e^{ i \theta nl} |\beta_n \rangle
% \ee

% \be
% |A_1^+ \rangle = \frac{1}{\sqrt{2}}(|\beta, 2, d_+ \rangle+|\beta, \bar 2, d_- \rangle)
% \ee

In Section~\ref{sec:CDW_clusters} of the main text we have described the construction of symmetrized cluster states, states which transform irreducibly under the crystal symmetry group ($D_{3d}$) of a single cluster. We gave explicit expressions for the states with $A_1^+ $ symmetry. Here we provide explicit expressions for the remaining states.

The two states with $A_2^+ $ symmetry are given by
\begin{align}
|A_2^+ (1) \rangle & = (|\beta, 2, d_+ \rangle-|\beta, \bar 2, d_- \rangle)/\sqrt{2}, \\
|A_2^+ (2) \rangle & = (|\gamma, 2, d_+ \rangle-|\gamma, \bar 2, d_- \rangle)/\sqrt{2}.
\end{align}
The three states with $A_1^- $ symmetry are given by
\begin{align}
|A_1^- ( 1 ) \rangle &  = |\beta, 3, d_0 \rangle, \\
|A_1^- ( 2 )\rangle & = (|\beta, 1, d_- \rangle+|\beta, \bar 1, d_+ \rangle)/\sqrt{2}, \\
|A_1^- ( 3 ) \rangle &= (|\gamma, 1, d_- \rangle-|\gamma, \bar 1, d_+ \rangle)/\sqrt{2}.
\end{align}
The three states with $A_2^- $ symmetry are given by
\begin{align}
|A_2^- ( 1 ) \rangle &  = |\gamma, 3, d_0 \rangle, \\
|A_2^- ( 2 )\rangle & = (|\beta, 1, d_- \rangle-|\beta, \bar 1, d_+ \rangle)/\sqrt{2}, \\
|A_2^- ( 3 ) \rangle &= (|\gamma, 1, d_- \rangle+|\gamma, \bar 1, d_+ \rangle)/\sqrt{2}.
\end{align}
There are fourteen doublets with $E^+ $ symmetry. Here we choose a basis for this doublet for which the threefold rotation is diagonal and only list one of the partners; the other may be obtained by applying the $\mathcal C_{2x}$ rotation. The seven state with $E^+ $ symmetry are then
\begin{align}
|E^+ ( 1 ) \rangle &  = |\alpha, 0, d_+ \rangle, \\
|E^+ ( 2 )\rangle & = |\beta, \bar 2, d_0 \rangle, \\
|E^+ ( 3 ) \rangle &= |\gamma, \bar 2, d_0 \rangle,\\
|E^+ ( 4 )\rangle & = |\beta, 0, d_+ \rangle, \\
|E^+ ( 5 ) \rangle &= |\gamma, 0, d_+ \rangle,\\
|E^+ ( 6 ) \rangle &= |\beta,  2, d_- \rangle, \\
|E^+ ( 7 ) \rangle &= |\gamma,  2, d_- \rangle.
\end{align}
Similarly, there are twelve doublets with $E^- $ symmetry. Listing only one of the two partners, we have
\begin{align}
|E^- ( 1 )\rangle & = |\beta, 1, d_0 \rangle, \\
|E^- ( 2 ) \rangle &= |\gamma, 1, d_0 \rangle,\\
|E^- ( 3 )\rangle & = |\beta, 3, d_+ \rangle, \\
|E^- ( 4 ) \rangle &= |\gamma, 3, d_+ \rangle,\\
|E^- ( 5 ) \rangle &= |\beta,  \bar 1, d_- \rangle, \\
|E^- ( 6 ) \rangle &= |\gamma,  \bar 1, d_- \rangle.
\end{align}

\end{document}